\documentclass[12pt]{article}
\usepackage{color} 
\usepackage{latexsym}
\usepackage{amssymb}
\usepackage{amsmath}
\usepackage[english]{babel}
\title{Comparison of quantization of charge transport in periodic and open
  pumps} 

\def\ds{\frac{\partial \psi_{nks}}{\partial s}}
\def\dk{\frac{\partial \psi_{nks}}{\partial k}}

\def\doublel{\langle \! \langle}
\def\doubler{\rangle \! \rangle}
\newcommand{\ep}[1]{\mathrm{e}^{#1}}

\begin{document}
\author{G.M. Graf and G. Ortelli\\
\normalsize\it
 Theoretische Physik, ETH-H\"onggerberg, 8093 Z\"urich, Switzerland}
\maketitle

\begin{abstract}

We compare the charges transported in two systems, a spatially periodic 
and an open
quantum pump, both depending periodically and adiabatically on time. The
charge transported in a cycle was computed by Thouless, respectively by 
B\"uttiker et al. in the two cases. We show that the
results agree in the limit where the two physical situations become the
same, i.e., that of a large open pump. 

\end{abstract}

\section{Introduction}
\label{sez:uno}

In this note we compare two systems, depending periodically and adiabatically
on time, which may exhibit quantized charge transport. We describe them in the
simplest possible situation. The first one, which may be called a periodic
quantum pump, is modelled as a 1-dimensional Fermi gas moving in a potential
which is periodic in space as well. Thouless \cite{Th} showed that charge
transport is quantized provided the Fermi energy remains in a gap throughout
the cycle. 
The second system is an open quantum pump and consists of a dot connected to
two leads, containing free Fermi gases. Particles impinging on the dot may be
transmitted through or reflected by it. The charge transported in a cycle has
been expressed by B\"uttiker et al. \cite{BTP} (see also \cite{B, ZSA}) in 
terms of the scattering matrix at Fermi energy. It is quantized in special 
cases only \cite{AK, SAA, AEGS}. At first sight
the descriptions of transport in periodic and open pumps may look unrelated,
because of the infinite, resp. finite extent of the two devices; or even
conflicting: in the first case transport is attributed to energies way below
the Fermi energy, which lies in a spectral gap; in the second the scattering
matrix matters only at Fermi energy. 

In order to compare the two approaches we consider the pump obtained by
truncating the potential of the periodic pump to finitely many periods, and
joining the ends to half-lines where particles move freely. In the limit where
the number of periods grows large we recover the original physical situation,
and one would wish the two approaches to agree on the result. This is shown in
this paper. 

A related comparison, though for finite dots, was made in \cite{C}, where the
scattering approach of \cite{BTP} was shown to agree with linear response
theory, an instance of which is the approach by Thouless. 

In Sections~\ref{sez:due} and \ref{sez:tre} we recall the results for periodic 
and open pumps respectively, to the extent needed for the comparison, which 
is made in Section~\ref{sez:quattro}.

\section{Transport in periodic pumps}
\label{sez:due}

We consider the Schr\"odinger Hamiltonian on the line $\mathbb{R}_x$
\begin{equation} \label{eq:uno}
  H(s) = -\frac{d^2}{dx^2} + V(x, s) \, ,
\end{equation}
where $V$ is doubly periodic: $V(x+ L,s) = V(x, s)$ and $V(x, s + T)= V(x, s)$.
Let $\psi_{nks}(x)$ be the solution of the time-independent Schr\"odinger
equation 
\begin{equation*}
  H(s) \psi_{nks} = E_{ns}(k) \psi_{nks} \, ,
\end{equation*}
with Bloch boundary condition 
\begin{equation*}
  \psi_{nks}(x + L) = \ep{\textrm{i}kL} \psi_{nks}(x) \, , \qquad (k \in
  \mathbb{R} \mod 2 \pi /L)
\end{equation*}
and normalized with respect to the inner product
\begin{equation*}
  \langle \phi, \psi \rangle = \frac{1}{L} \int_0^L dx \, \overline{\phi} (x)
  \psi (x) \, .
\end{equation*}
The index $n = 1,2, \, ..$ labels the bands. The Fermi energy is assumed to
lie in a spectral gap of $H(s)$ for all $s$. 

Thouless discusses the evolution of the Fermi sea under the non-autono-\\mous 
Hamiltonian $H(\omega t)$
in the adiabatic limit $\omega \rightarrow 0$. The charge transported through
$x=0$ in a cycle of period $T/\omega$ is found to be  
\begin{equation*}
  C = {\sum_n}^*
 \frac{\textrm{i}}{2 \pi} \int_0^T \!\! ds
  \int_0^{2 \pi / L} \!\! dk \Big( \langle \ds , \dk \rangle - \langle \dk
  , \ds \rangle \Big) \, ,   
\end{equation*}
where the star indicates that the sum extends over filled bands $n$ only. Each
of its terms is an integer defining the Chern number of the $U(1)$ fiber
bundle $\psi_{nks}$ over the torus $\mathbb{T} = \mathbb{R}^2
/ (T, 2 \pi / L)$. That number reflects the obstruction to choosing the
phase of $\psi _{nks}$ in a continuous way on all of $\mathbb{T}$.
  
Thouless provides the following alternate definition of $C$, related to the
above by analytic continuation. At energies in a gap, and specifically at the
Fermi energy, the solutions of the time-independent Schr\"odinger equation are
unbounded, and two linearly independent ones may be picked by the condition
\begin{equation}\label{eq:due}
  \psi_{\pm ,s} (x+L) =   (-1)^n \ep{\pm \kappa x}\psi_{\pm ,s}(x)
\end{equation}
for some $\kappa = \kappa (s) > 0$, where $n$ refers to the gap following the
$n$-th band. The functions $\psi_{\pm}$ may be assumed real. Then $C$ equals 
the number of nodes of $\psi_-$ traversing a reference point, say $x = 0$, 
as $s$ completes a cycle. A node contributes positively if it runs from left 
to right.

\section{Transport in open pumps}
\label{sez:tre}

We consider the Hamiltonian (\ref{eq:uno}) where $V(x, s)$ is now of
compact support in $x$, but still periodic in $s$. The autonomous dynamics it
generates for a fixed value of $s$ yields a scattering matrix 
\begin{equation*}
  S(E, s) = \big( S_{ij} \big)_{i,j=1}^2 = \left( \begin{array}{cc} r & t'
      \\ t & r' \end{array} \right) \, , 
\end{equation*}
where the entry $S_{ij}$ is the amplitude for a particle of energy $E$,
incident from lead $j$ to be scattered into lead $i$; or, more explicitly,
$t$, $r$ (resp. $t'$, $r'$) are the transmission and reflection amplitudes for
a wave incident from the left (resp. right). We shall henceforth set $E=E_F$
and drop it from the notation. 

B\"uttiker et al. \cite{BTP} investigate the motion
of particles governed by the non-autonomous Hamiltonian $H(\omega t)$, again
in the adiabatic limit. The charge delivered to lead $j$ in a cycle is
\begin{equation*}
  \langle Q_j \rangle = \frac{\textrm{i}}{2 \pi} \int_{s=0}^{s=T} ((dS) \,
  S^*)_{jj} \, , 
\end{equation*}
where $\langle \cdot \rangle$ denotes a quantum mechanical expectation
value. For the same situation Ivanov et al. \cite{ILL} (following \cite{LL}) 
computed the variance of the same quantity
\begin{align*}
  \doublel  Q_j^2 \doubler  &= \frac{1}{(2 \pi)^2}
    \int_{-\infty}^{\infty} \int_0^T ds \, ds' \, \frac{1-|\big(S(s)S^* (s')
      \big)_{jj}|^2}{(s-s')^2} \\
      &= T^2 \int_0^T \int_0^T ds \, ds' \, \frac{1-|\big( S(s)S^* (s')
      \big)_{jj}|^2}{\sin^2 \frac{2 \pi}{T}(s-s')} 
\end{align*}
(note that denominator and numerator both vanish quadratically at $s=s'$). For
the left lead ($j=1$) this reads
\begin{align}\label{eq:tre}
  \langle Q_1 \rangle &= \frac{\textrm{i}}{2 \pi} \int_{s=0}^{s=T} \big(
  \overline{r} 
  dr + \overline{t}' dt' \big) \, , \\
  \doublel Q_1^2 \doubler &= T^2 \int_0^T \int_0^T ds \, ds' \,
  \frac{1-|\big( r(s) \overline{r} (s') + t'(s) \overline{t}'(s')
    \big)|^2}{\sin^2 \frac{2 \pi}{T}(s-s')} \, . \nonumber
\end{align}
It has been noticed \cite{AEGS} that if the $j$-th row of $S(s)$ changes with
$s$ by multiplication with a phase $u(s)$ ($|u(s)| = 1$), then

\begin{equation}\label{eq:quattro}
  \langle Q_j \rangle = \frac{\textrm{i}}{2 \pi} \int_{s=0}^{s=T} \overline{u}
  \, du 
\end{equation}
is the negative of the winding number of $u$, and 
$\doublel Q_j^2 \doubler = 0$,
meaning that the charge transport is quantized. For the left lead that
condition amounts to $z = r/t' \in \mathbb{C} \cup \{ \infty \}$ remaining put
as a point on the Riemann sphere.

\section{The comparison}
\label{sez:quattro}

We compare the periodic pump to the open one obtained from it by truncating
the potential to $N$ periods:
\begin{equation*}
  H(s) = -\frac{d^2}{dx^2} + \chi_{\lbrack 0, N L \rbrack } (x) V(x, s)
    \, . 
\end{equation*}
Let us determine the scattering matrix at the Fermi energy. For a wave
incident from the left the solution is of the form
\begin{equation*}
   \left\{ \begin{array}{lll} \ep{\textrm{i} p x} + r_N \ep{-\textrm{i}px} &,&
       (x \leq 0) \\  
  A_+ \psi_{+}(x) + A_- \psi_{-}(x)  &,&  (0 \leq x \leq N L) \\ 
  t_N \ep{\textrm{i} p x} &,&  (x \geq N L) \end{array} \right.
\end{equation*}
with $p=\sqrt{E_F}$, $\psi_\pm$ as specified in (\ref{eq:due}), and $s$
temporarily omitted from the notation. Within the barrier the Wronskian of
this solution and $\psi_+$ (or $\psi_-$) is constant, and in particular equal
at $x=0$ and at $x=NL$. The matching conditions thus amount to
\begin{equation*}
  W (\ep{\textrm{i}px} + r_N \ep{-\textrm{i}px}, \psi_{\pm})|_{x=0} = W ( t_N
  \ep{\textrm{i} p x}, 
  \psi_{\pm})|_{x=N L} \, ,
\end{equation*}
where $W (\phi, \psi)|_x = \phi (x) \psi' (x) - \phi' (x) \psi (x)$. 
Setting $W_\pm = W (\ep{\textrm{i}px}, \psi_\pm) |_{x=0}$ and using 
\begin{equation*}
  W (\ep{-\textrm{i}px}, \psi_\pm )|_{x=0} = \overline{W}_\pm \, , \qquad W
  (\ep{\textrm{i}px} \, ,
  \psi_\pm )|_{x = N L} = 
(-1)^{nN} \ep{\pm \kappa NL} \ep{\textrm{i}pNL} W_\pm \, ,
\end{equation*}
we find
\begin{align*}
  r_N &= -u_- \frac{1 - \ep{-2 \kappa NL}}{1 - \ep{-2 \kappa NL} u_- u_+^{-1}}
  \, ,  \\
  t_N &= (-1)^{nN} \ep{- \kappa NL}\ep{-\textrm{i}pNL} \frac{1 - u_-
    u_+^{-1}}{1 - \ep{-2 \kappa NL} u_- u_+^{-1}}
\end{align*}
with
\begin{equation}\label{eq:cinque}
  u_\pm = \frac{W_\pm}{\overline{W}_\pm} = \frac{\psi_\pm ' (0) - \textrm{i}p
    \psi_\pm 
    (0)}{\psi_\pm ' (0) + \textrm{i}p \psi_\pm (0)} \, .
\end{equation}
Owing to the invariance of the Hamiltonian under time reversal, 
$S$ is symmetric,
i.e., $t_N' = t_N$. In the limit of a long barrier we have
\begin{equation*}
  r = \lim_{N \rightarrow \infty} r_N = -u_- \, , \qquad
  t' = \lim_{N \rightarrow \infty} t_N ' = 0 \, ,
\end{equation*}
and the condition for quantized transport is attained exponentially fast in
$N$. Restoring the dependence on $s$, the charge (\ref{eq:tre}) or
(\ref{eq:quattro}) delivered to the left lead becomes

\begin{equation*}
  \lim_{N \rightarrow \infty} \langle Q_1 \rangle = \frac{\textrm{i}}{2 \pi}
  \int_{s=0}^{s=T} \big( \overline{r} dr + \overline{t}' dt' \big) =
  \frac{\textrm{i}}{2 \pi} \int_{s=0}^{s=T} \overline{u}_- du_- \, ,
\end{equation*}
which, up to the sign, is the winding number of the phase $u_- (s)$. The
charge crossing $x=0$ in the positive direction is thus given by the winding
number itself. 

Finally, we show that this result agrees with the Chern number of Thouless, 
as characterized at the end of Section~\ref{sez:due}. Whenever a node
of $\psi_{-,s}$ crosses $x=0$ from the left, $\partial \psi_-/\partial
  s|_{x=0}$ and $\partial \psi_-/\partial x|_{x=0}$ have opposite
signs. Hence $u_- (s)$, which moves along the unit circle, see
(\ref{eq:cinque}), crosses $u=1$ from below, counting $+1$ to its winding
number; nodes crossing $x=0$ from the right contribute $-1$.\\
  
\noindent
{\bf Acknowledgements.} We thank Y. Avron for useful discussions.

\end{document}